\documentclass[sigconf,9pt]{acmart}
\usepackage{booktabs}
\usepackage{multirow}
\usepackage{tikz}
\usepackage{amsmath}
\usepackage{siunitx}
\usepackage[abbreviations]{foreign}
\usepackage{xspace}
\usepackage[inline]{enumitem}
\usepackage[algo2e,ruled,lined,boxed,commentsnumbered, noend, linesnumbered, procnumbered]{algorithm2e}
\definecolor{orange}{rgb}{1,0.5,0}
\definecolor{Green}{rgb}{0,0.5,0}
\definecolor{Blue}{rgb}{0,0,1}
\usepackage{svg}
\usepackage{pifont}%

\usepackage{acronym}
\acrodef{DL}{decentralized learning}
\acrodef{ML}{machine learning}
\acrodef{D-PSGD}{decentralized parallel stochastic gradient descent}
\acrodef{FL}{federated learning}
\acrodef{SGD}{stochastic gradient descent}
\acrodef{IID}{independent and identically distributed}
\acrodef{non-IID}{non independent and identically distributed}
\acrodef{RMSE}{root mean square error}
\acrodef{RMW}{random model walk}
\acrodef{GL}{gossip learning}
\acrodef{EL}{epidemic learning}
\acrodef{DWT}{discrete wavelet transform}
\acrodef{FFT}{fast Fourier transform}
\acrodef{MI}{mutual information}
\acrodef{DP}{differential privacy}
\acrodef{VN}{virtual node}
\acrodef{RN}{real node}
\acrodef{LDP}{local differential privacy}
\acrodef{PNDP}{pairwise network differential privacy}
\acrodef{PNLDP}{pairwise network local differential privacy}
\acrodef{GI}{gradient inversion}
\acrodef{CML}{collaborative machine learning}
\acrodef{TPR}{true positive rate}
\acrodef{FPR}{false positive rate}
\acrodef{LLM}{large language model}
\acrodef{RAG}{retrieval-augmented generation}
\acrodef{ANN}{approximate nearest neighbor}
\acrodef{LRU}{least-recently used}
\acrodef{FIFO}{first-in, first-out}
\acrodef{MMLU}{Massive Multitask Language Understanding}
\acrodef{NNS}{nearest neighbor search}
\acrodef{LSH}{locality-sensitive hashing}
\acrodef{TTFT}{Time To First Token}
\acrodef{HNSW}{hierarchical navigable small world} %
\newcommand{\sys}{\textsc{Proximity}\xspace}

\newcommand{\mmlu}{\ac{MMLU}\xspace}
\newcommand{\medrag}{\textsc{MedRAG}\xspace}
\newcommand{\tripclick}{\textsc{TripClick}\xspace}
\newcommand{\pubmed}{\textsc{PubMed}\xspace}
\newcommand{\diskann}{\textsc{DiskANN}\xspace} %
\graphicspath{ {figures/} }
\usepackage{tikz}
\usepackage[eulergreek]{sansmath}
\usepackage{pgfplots}
\usepackage{pgfplotstable}
\usepackage{cleveref}
\usepackage{comment}
\crefname{assumption}{assumption}{assumptions}

\pgfplotsset{compat=newest}
\usepgfplotslibrary{external,units,colorbrewer,groupplots,fillbetween,statistics}
\tikzsetexternalprefix{figures/}
\tikzset{external/mode=list and make}
\usetikzlibrary{patterns,shapes.misc}

\def\overleafhome{/tmp}
\newcommand{\inputplot}[2]{%
	\ifx\homepath\overleafhome%
	\IfBeginWith{#1}{plots}{\includegraphics{main-figure#2.pdf}}{#1}%
	\else%
	{\sffamily\scriptsize\input{#1}}
	\fi
}

\newcommand{\newgroupwidth}[2]%
{\expandafter\xdef\csname groupwidth#1\endcsname{#2}}

\newcounter{groupwidth}
\newsavebox{\groupwidthbox}
\makeatletter
{\edef\groupnumber{#1}%
	\stepcounter{groupwidth}%
	\@ifundefined{groupwidth\thegroupwidth}{\pgfmathsetlengthmacro{\mywidth}{\linewidth/\groupnumber}}%
	{\expandafter\let\expandafter\mywidth\csname groupwidth\thegroupwidth\endcsname}%
	\begin{lrbox}{\groupwidthbox}%
		\tikzset{/pgfplots/width={\mywidth}}%
		\ignorespaces}%
	{\end{lrbox}%
	\usebox\groupwidthbox
	\pgfmathsetlengthmacro{\mywidth}{\mywidth + (\linewidth - \wd\groupwidthbox)/\groupnumber}
	\immediate\write\@auxout{\string\newgroupwidth{\thegroupwidth}{\mywidth}}}
\makeatother
\usepackage{amsmath,amssymb,amsfonts,amsthm}
\usepackage{enumitem}
\usepackage{thm-restate}
\usepackage{bbm}
\theoremstyle{definition}

\theoremstyle{remark}

\allowdisplaybreaks

\copyrightyear{2025}
\acmYear{2025}
\setcopyright{acmlicensed}\acmConference[MIDDLEWARE '25]{26th International Middleware Conference}{December 15--19, 2025}{Nashville, TN, USA}
\acmBooktitle{26th International Middleware Conference (MIDDLEWARE '25), December 15--19, 2025, Nashville, TN, USA}
\acmDOI{10.1145/3721462.3770776}
\acmISBN{979-8-4007-1554-9/2025/12}

\setcopyright{acmlicensed}

\makeatletter
\AtBeginDocument{%
	\def\ltx@label#1{\cref@label{#1}}%
	
	\def\label@in@display@noarg#1{\cref@old@label@in@display{#1}}%
	
	\def\label@in@mmeasure@noarg#1{%
		\begingroup
		\measuring@false
		\cref@old@label@in@display{#1}%
		\endgroup
	}%
}
\makeatother
 
\begin{document}


\title{Leveraging Approximate Caching for Faster Retrieval-Augmented Generation}

\author{Shai Bergman}
\affiliation{
  \institution{Huawei Research}
  \city{Zurich}
  \country{Switzerland}
}

\author{Anne-Marie Kermarrec}
\affiliation{
  \institution{EPFL}
  \city{Lausanne}
  \country{Switzerland}
}

\author{Diana Petrescu}
\affiliation{
  \institution{EPFL}
  \city{Lausanne}
  \country{Switzerland}
}

\author{Rafael Pires}
\affiliation{
  \institution{EPFL}
  \city{Lausanne}
  \country{Switzerland}
}

\author{Mathis Randl}
\affiliation{
  \institution{EPFL}
  \city{Lausanne}
  \country{Switzerland}
}
\authornote{Corresponding author}

\author{Martijn de Vos}
\affiliation{
  \institution{EPFL}
  \city{Lausanne}
  \country{Switzerland}
}

\author{Ji Zhang}
\affiliation{
  \institution{Huawei Research}
  \city{Zurich}
  \country{Switzerland}
}

\renewcommand{\shortauthors}{Bergman et al.}

\begin{abstract}

Retrieval-augmented generation (RAG) improves the reliability of large language model (LLM) answers by integrating external knowledge.
However, RAG increases the end-to-end inference time since looking for relevant documents from large vector databases is computationally expensive.
To address this, we introduce \textsc{Proximity}, an approximate key-value cache that optimizes the RAG workflow by leveraging similarities in user queries.
Instead of treating each query independently, \textsc{Proximity} reuses previously retrieved documents when similar queries appear, substantially reducing the reliance on expensive vector database lookups.
To efficiently scale, \textsc{Proximity} employs a locality-sensitive hashing (LSH) scheme that enables fast cache lookups while preserving retrieval accuracy.
We evaluate \textsc{Proximity} using the \textsc{MMLU} and \textsc{MedRAG} question-answering benchmarks.
Our experiments demonstrate that \textsc{Proximity} with our LSH scheme and a realistically-skewed \textsc{MedRAG} workload reduces database calls by 77.2\% while maintaining database recall and test accuracy.
We experiment with different similarity tolerances and cache capacities, and show that the time spent within the \textsc{Proximity} cache remains low and constant (\SI{4.8}{\micro\second}) even as the cache grows substantially in size.
Our results demonstrate that approximate caching is a practical and effective strategy for optimizing RAG-based systems.

\end{abstract}


\maketitle

\section{Introduction}
\label{sec:intro}
\Acfp{LLM} have revolutionized natural language processing by demonstrating strong capabilities in tasks such as text generation, translation, and summarization~\cite{kaplan2020scaling}.
Despite their increasing adoption, a fundamental challenge is to ensure the reliability of their generated responses~\cite{zhou2024larger,huang2025survey}.
A particular issue is that \Acp{LLM} are prone to \emph{hallucinations} where they confidently generate false or misleading information, which limits their applicability in high-stake domains such as healthcare~\cite{ji2023towards} and finance~\cite{roychowdhury2024journey}.
Moreover, their responses can be inconsistent across queries, especially in complex or specialized domains, making it difficult to trust their outputs without extensive verification by domain experts~\cite{zhou2024larger,lee2024one}.

\Acf{RAG} is a popular approach to improve the reliability of \ac{LLM} answers~\cite{lewis2020retrieval}.
\Ac{RAG} combines the strengths of neural network-based text generation with external information retrieval.
This technique first retrieves relevant documents from an external database based on the user query and includes them in the \ac{LLM} prompt before generating a response.
Both user queries and documents are often represented as high-dimensional embedding vectors that capture semantic meanings, and these embeddings are stored in a vector database.
Retrieving relevant documents involves finding embeddings in the database that are the closest to the query embedding, a process known as \ac{NNS}.
Thus, \ac{RAG} enables the \ac{LLM} to use reliable sources of information without the need to modify the model parameters through retraining or fine-tuning~\cite{csakar2025maximizing}.

At the same time, the \ac{NNS} operation that is part of the \ac{RAG} workflow becomes computationally expensive for large vector databases~\cite{shen2024towards,zhu2024accelerating}.
Thus, \ac{RAG} can significantly prolong the inference end-to-end time~\cite{jiang2024piperag,accelerating_rag_asplos}.
To mitigate the latency increase of \ac{NNS}, we observe that user query patterns to conversational agents or search engines often exhibit spatial and temporal locality, where specific topics may experience heightened interest within a short time span~\cite{frieder2024caching} or otherwise exhibit strong bias towards some queries~\cite{meats2007using}.
Similar queries are likely to require and benefit from the same set of retrieved documents in such cases, even if they are not exactly syntactically equal.
Building on this observation, we reduce the database load by caching and reusing results from similar past user queries.
This approach contrasts with conventional \ac{RAG} systems~\cite{zhu2024accelerating,ray2024ragserve} that treat each query as independent from the others without exploiting access patterns.
However, exact embedding matching is ineffective when queries are phrased differently, since even slight rephrasings of a query typically yield different embedding vectors.
To this end, we introduce a novel \emph{approximate caching} mechanism that uses the semantic information in the embeddings of queries, computing similarity in the embedding space and incorporating a similarity threshold to address the exact matching problem.
Approximate caching allows for some level of tolerance when determining relevant cache entries.

This work introduces \sys, a novel approximate key-value cache specifically designed for \ac{RAG}-based \ac{LLM} systems.
By intercepting queries before they reach the vector database and by leveraging previously retrieved results for similar queries, \sys reduces the computational cost of \ac{NNS} and minimizes database accesses, effectively lowering the total end-to-end inference latency of the \ac{RAG} pipeline.
Specifically, we store past document queries in an approximate key-value cache,
where each key corresponds to the embedding of a previous query, and the associated value is the set of relevant documents retrieved for that query.
When a new query is received, the cache checks if there is a cache entry within some similarity threshold $ \tau $ and if so, returns the entry closest to the incoming query.
On a cache hit, the corresponding documents are returned, bypassing the need for a database lookup.
In case of a cache miss, the system queries the vector database to retrieve relevant documents for the new query.
The cache is then updated with the new query and retrieved documents, and the \ac{RAG} pipeline proceeds as usual.
However, to determine whether previously cached queries are sufficiently close to an incoming query, we need to do a linear scan over all cached entries, which becomes computationally expensive as the size of the cache grows.
To address this scalability concern, we leverage a \acf{LSH} scheme and introduce \sys-LSH, a variant of our approximate cache.

We implement \sys and evaluate our system using the \mmlu~\cite{hendrycks2020measuring}, and \medrag~\cite{xiong-etal-2024-benchmarking} benchmarks, which are commonly used to evaluate \ac{RAG} frameworks. We provide two versions of the \medrag benchmarks, one where queries are repeated four times each in slight variations, and one where queries are repeated according to a Zipfian distribution with parameter 0.8.
Compared to a \ac{RAG} pipeline without caching, \sys brings significant speed improvements while maintaining retrieval accuracy.
Specifically, we find that \sys reduces the latency of document retrieval by up to 59\% for \mmlu and 75\% for \medrag with no or only a marginal decrease in accuracy.
These findings hold across all versions of the experiments, with and without bias towards hot queries.
\textsc{Proximity} with our \ac{LSH} scheme and a realistically skewed \textsc{MedRAG} workload reduces database calls by 77.2\% and reduces document retrieval latency by 72.5\% while maintaining test accuracy.
These results highlight the viability and effectiveness of using approximate caching to improve the speed of \ac{RAG}-based \ac{LLM} systems.
To the best of our knowledge, \sys is the first system to apply approximate caching to reduce the document retrieval latency in \ac{RAG} pipelines during \Acp{LLM} inference.

\hyphenation{data-sets}
Our contributions are as follows:
\begin{itemize}
    \item We introduce \sys-FLAT, a novel approximate key-value cache for RAG pipelines that leverages spatial and temporal similarities in user queries to reduce the overhead of document retrieval (\Cref{sec:design}).
    \sys-FLAT includes a similarity-based caching strategy that significantly reduces retrieval latency while maintaining high response quality.
    \item We enhance the scalability of \sys-FLAT by introducing \sys-LSH, an efficient and scalable variant that relies on \acf{LSH} to determine sufficiently similar cache entries (\Cref{sec:proximity_lsh}).
    \item We benchmark both cache variants using two standard datasets, demonstrating substantial improvements in cache hit rates and query latency while maintaining comparable test accuracies (\Cref{sec:evaluation}).
    These improvements are quantified relative to a baseline \ac{RAG} pipeline without caching.
    We also analyze the impact of the cache capacity and similarity tolerance, providing insight into optimizing retrieval performance for workloads with differing characteristics.
\end{itemize}

\section{Background and motivation}
\label{sec:prelims}

\begin{figure}
    \centering
    \includegraphics{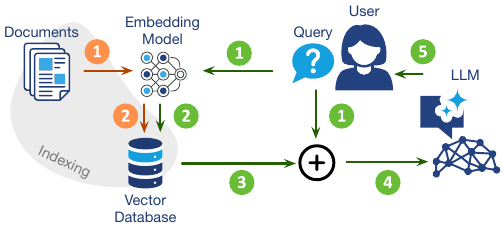}
    \caption{The \ac{RAG} workflow.}
    \Description[This image describes the RAG workflow.]{Documents are stored in a vector database. When the user emits a query, it is also sent to the database so that the most relevant documents are appended to the query before being sent to the LLM.}
    \label{fig:rag}
\end{figure}

We first detail the \ac{RAG} workflow in \Cref{sec:rag_workflow} and then outline the process to retrieve the documents relevant to a user query from the vector database in \Cref{sec:rag_vector_search}.
We then show in \Cref{sec:skew} the existence of skew in search engine queries, which provides ground for caching opportunities.

\subsection{Retrieval-augmented generation (RAG)}
\label{sec:rag_workflow}
\Acf{RAG} is a technique that enhances the capabilities of \acp{LLM} by integrating information retrieval before the generation process~\cite{lewis2020retrieval}.
\ac{RAG} typically enables higher accuracy in benchmarks with a factual ground truth, such as multiple-choice question answering~\cite{dayarathne2024comparing}.

\Cref{fig:rag} shows the \ac{RAG} workflow that consists of the following eight steps.
Before \ac{LLM} deployment, documents are first converted into high-dimensional embedding vectors using an embedding model {\color{orange}\ding{202}} and stored in a vector database {\color{orange}\ding{203}}.
Optionally, documents are divided into smaller chunks before embedding, which improves retrieval accuracy by allowing the system to retrieve the most relevant sections of a document rather than entire documents.
These two steps comprise the indexing phase.
When the user sends a query to the \ac{LLM} {\color{Green}\ding{202}}, this query is first converted to an embedding vector {\color{Green}\ding{203}} using the same embedding model as used for the indexing and passed to the retriever.
The vector database then searches for embeddings close to the query embedding using some distance metric and returns the relevant documents related to this embedding  {\color{Green}\ding{204}}.
These documents and the user query are combined into a single prompt and passed to the \ac{LLM} {\color{Green}\ding{205}}.
The \ac{LLM} response is then returned to the user.

\subsection{Vector search}
\label{sec:rag_vector_search}
The vector search during the \ac{RAG} workflow obtains relevant embeddings from a vector database based on the embedding vector of the user query.
Vector databases are databases that potentially store a vast amount of $n$-dimensional real-valued vectors and are optimized to solve the \acf{NNS}, \ie, finding the $k$ elements contained in the database that are the closest to a given query~\cite{pan2024survey}.
The similarity metric to be minimized is typically L2, cosine, or inner-product, and is fixed before deployment.
This lookup returns a ranked list of indices corresponding to resulting embeddings, and these indices can then be used to obtain the documents that will be sent along with the user prompt to the \ac{LLM}.

Due to the high dimensionality of embeddings and the sheer volume of data in modern vector databases~\cite{borgeaud2022improving}, performing vector searches at scale poses significant computational challenges.
\ac{NNS} requires comparing query embeddings with millions or billions of stored vectors, which becomes slow and computationally expensive as the database grows~\cite{chen2021spann}.
Even with optimized index structures such as as \ac{HNSW}~\cite{malkov2018efficient} or quantization-based approaches~\cite{jegou2010product}, to maintain low-latency retrieval while ensuring high recall remains difficult.

Several studies show that vector search can account for a significant portion of the end-to-end latency in \ac{RAG}-based \ac{LLM} systems~\cite{rago, shen2024towards, quinnasplos25}.
In particular, Shen et al.~\cite{shen2024towards} report that the average \ac{TTFT} increases from \SI{495}{\milli\second} to \SI{965}{\milli\second} after deploying \ac{RAG}, with a significant share of this overhead (71.8\%) attributed to the vector database lookup.
\ac{TTFT} captures the latency between sending a query and receiving the first output token from the \ac{LLM}.
The remaining increase comes from a slightly longer \ac{LLM} pre-fill stage caused by processing the additional retrieved documents.
We note that the proportion of \ac{TTFT} required for vector search depends on factors such as the \ac{LLM} size, the vector database implementation, and the number of retrieved vectors.
Nevertheless, these findings highlight that vector search latency can become a serious bottleneck in \ac{RAG} systems.

\subsection{Skew in user queries}
\label{sec:skew}
RAG pipelines often face heavily skewed user query distributions, similar to those found in search engines and conversational agents \cite{spink2001searching,teevan2007information,srivatsa2024preble,gim2024prompt,hu2024epic}. Specifically, a few popular queries dominate, while most occur rarely or only once. Moreover, \Acp{LLM} are increasingly being integrated in search engine ecosystems, \eg, ChatGPT Search \cite{chatgpt_search}, Google AI Overviews~\cite{google_ai_overviews} and Microsoft Copilot Search~\cite{microsoft_copilot_search}.
These systems typically rely on \ac{RAG} pipelines to ground responses in knowledge sources.
In such systems, the documented skew and locality properties of search queries \cite{spink2001searching,teevan2007information} carry over to \ac{LLM}-based retrieval, making caching particularly appealing.

To quantify this skew, we analyze the \tripclick dataset, a large collection of user interactions in a health-focused search engine.
\tripclick comprises approximately 5.2 million user interactions collected from the Trip Database~\cite{meats2007using} between 2013 and 2020.
The dataset includes around \num{700000} unique free-text queries and \num{1.3} million query-document relevance pairs, making it a valuable resource for studying user search behavior.

\begin{figure}[t]
    \centering
    \includegraphics{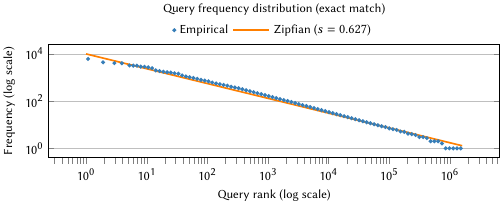}
    \caption{The TripClick empirical query frequencies, along with the expected frequencies from the matching Zipfian distribution.
    }
    \Description[todo.]{todo}
    \label{fig:tripclick}
\end{figure}

\Cref{fig:tripclick} shows the exact-match query frequency distribution with the query rank on the horizontal axis and frequency of the query on the vertical axis (both axis in log scale).
The empirical distribution of query frequencies closely matches a Zipfian curve with an exponent of $\approx 0.627$.
This power law distribution is not unique to the \tripclick dataset.
In fact, several studies show that user queries in other domains also often follow a Zipfian power law distribution, with exponent $ s $ roughly between $s = 0.6 $ and $ s = 2.5$, resulting in a strong bias towards a few topics of interest~\cite{falchi:2008:similaritycache,sabnis:2023:grades,frieder2024caching}.
This bias is typical of natural language and brings opportunity for \emph{caching}: the most frequent queries and their semantic variants generate a disproportionate share of retrieval workload and their results may be reused, therefore reducing computational load~\cite{baeza2007impact}.

\begin{figure}
    \centering
    \includegraphics{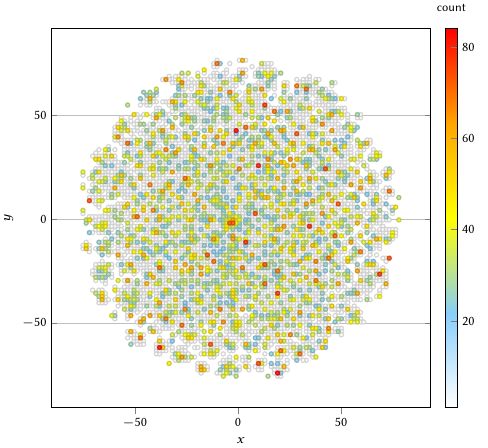}
    \caption{The two-dimensional projection of queries in the \tripclick dataset, each one encoded using the \textsc{MedCPT} embedding model.}
    \Description[A cloud of points that shows visible clusters in the embeddings of TripClick]{A cloud of points that shows visible clusters in the embeddings of TripClick}
    \label{fig:tripclick_embeddings}
\end{figure}

Beyond exact query duplicates, many queries differ only slightly in their written form but convey highly similar intent.
These include rephrasings, minor spelling variations, synonym substitutions, or changes in the word order (\eg, “best treatment for asthma” vs. “asthma best therapies”).
While these variations are not captured by exact matching, they often lie close together in a (latent) embedding space.
To analyze this effect, we embed the queries in the \tripclick dataset using the \textsc{MedCPT}~\cite{jin2023medcpt} embedding model and project the resulting vectors (which have a dimension of 768) into two dimensions using a combination of principal component analysis (PCA)~\cite{shlens2014tutorialprincipalcomponentanalysis} as a preprocessing step, followed by t-distributed stochastic neighbor embedding (t-SNE)~\cite{van2008visualizing} for visualization purposes.
We show the two-dimensional embedding space in \Cref{fig:tripclick_embeddings} as a 100x100 grid.
For each point in the grid we count the number of embeddings in it and color it accordingly.
This figure highlights that many queries cluster based on semantic content, even when their wording differs.

Thus, we empirically demonstrate using the \tripclick dataset that
\begin{enumerate*}[label=\emph{(\roman*)}]
\item user queries frequently repeat exactly, following a power‐law distribution; and
\item that syntactically different queries cluster in the embedding space, indicating semantic relationships between distinct queries.
\end{enumerate*}
These two findings are relevant for \ac{RAG} systems, which leverage the embedding space.
In such systems, each user query is mapped to a high-dimensional vector, and the nearest neighbors in the document index are retrieved based on this vector.
When two queries are semantically similar, not only are their embeddings close together, but the resulting retrieved documents often overlap.
In other words, different queries may map to the same or similar document sets.
\emph{This redundancy means that previously retrieved documents can be cached and reused for future queries with similar embeddings, reducing the number of expensive \ac{NNS} operations in the vector database.}

\section{Design of \sys}
\label{sec:design}

Guided by the above insights, we design a caching mechanism that reduces the need for repeated \acp{NNS} by reusing previously retrieved documents for similar queries.
More specifically, we leverage \emph{approximate caching} to accelerate \ac{RAG} document retrieval.
Even if the documents retrieved for a given query are not the most optimal results that would have been obtained from a full database lookup, they can still provide valuable context and relevant information for the \ac{LLM}, allowing the system to maintain good accuracy while reducing retrieval latency.
Among others, our approximate cache is parameterized with a similarity threshold $\tau$.
If the distance between two query embeddings $ q_1 $ and $ q_2 $ is equal to or less than $ \tau $, we consider these embeddings alike and return similar documents if they are available in the cache.
Cache hits thus bypass more expensive vector database searches.
However, to determine whether previous cached queries are close to an incoming query, we need to do a linear scan over all cached entries, which becomes computationally expensive as the size of the cache grows.
Therefore, the two technical challenges are
\begin{enumerate*}[label=\emph{(\roman*)}]
\item defining an effective set of hyperparameters, such as the similarity threshold that maximizes cache hits without compromising response relevance, as well as the optimal cache size that enables high cache coverage while still being computationally attractive, and
\item ensuring low computational overhead of each query lookup as the number of cached entries grows.
\end{enumerate*}

We now present the design of \sys, an approximate key-value cache designed to accelerate \ac{RAG} pipelines.
\sys is agnostic of the specific vector database being used but assumes that this database has a \textsc{retrieveDocumentIndices} function that takes as input a query embedding and returns a sorted list of document indices whose embeddings are close to the query embedding.
In \Cref{sec:retrieving_vectors} we first present the high-level process overview of retrieving vectors with \sys.
Then in \Cref{sec:proximity_lsh} we introduce \acf{LSH} in the context of approximate caching, one of the key optimizations implemented to reduce the latency of a search in the \sys cache.
Finally, we elaborate in \Cref{sec:cache_design} on the parameterization and components of our caching mechanism.

\begin{algorithm2e}[t]
	\caption{Cache lookup in \sys-FLAT}
	\label{algo:search}

	\DontPrintSemicolon
	\SetKwProg{Fn}{Procedure}{:}{}
	\SetKwInOut{Parameter}{Input}
	\SetKwInOut{Constructor}{Stored state}
	\SetKwInOut{Output}{Output}

	\Constructor{\quad similarity tolerance $\tau$, cache capacity $ c $,\\\quad vector database $\mathcal{D}$, key-value dict $\mathcal{C}$ = \{\}}

	\Fn{\textsc{lookup}($q$)}{
		$dists = [(k, \textsc{distance}(q, k)) \text{ for } k \text{ in } \mathcal{C}\text{.keys}]$
		\label{algo1computedists}

		$(key, min\_dist) \gets \text{min\_by\_dist}(dists)$
		\label{algo1minby}

		\If{$ min\_dist \leq \tau$}{
			\Return $\mathcal{C}[key]$
			\label{algo1earlyreturn}
		}

		$\mathcal{I} \gets \mathcal{D}.\textsc{retrieveDocumentIndices}(q)$
		\label{algo1dbcall}

		\If{$|\mathcal{C}| \geq c$}{
			\label{algo1ifevict}
			$\mathcal{C}$.\textsc{evictOneEntry}()
		}

		$\mathcal{C}[q] \gets \mathcal{I}$
		\label{algo1updatecache}

		\Return $\mathcal{I}$
		\label{algo1returnmiss}
	}
\end{algorithm2e}

\begin{figure}
    \centering
    \includegraphics[width=\columnwidth]{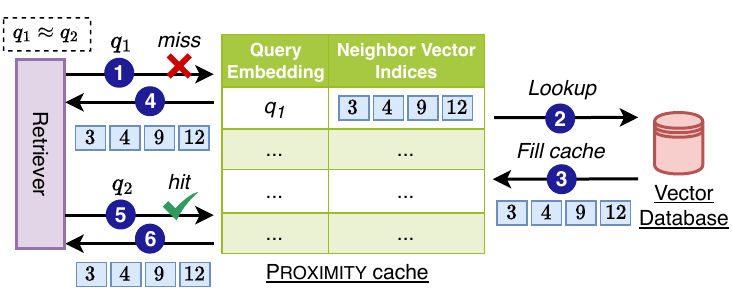}
    \caption{The design and workflow of the \sys approximate cache when receiving two subsequent, similar query embeddings $ q_1 $ and $ q_2 $. $ q_1 $ results in a cache miss whereas $ q_2 $ results in a hit, returning similar document indices as for $ q_1 $.}
    \label{fig:cache_design}
    \Description[When a cache entry doesn't match anything, it gets added to the cache after fetching the correct contents. If it does match something, even approximately, the previously fetched elements are returned.]{When a cache entry doesn't match anything, it gets added to the cache after fetching the correct contents. If it does match something, even approximately, the previously fetched elements are returned.}
\end{figure}

\subsection{Retrieving relevant documents with \sys-FLAT}
\label{sec:retrieving_vectors}

We first present the basic version of our cache, named \sys-FLAT, and later describe an optimized version of our cache in \Cref{sec:proximity_lsh}.
\Cref{algo:search} shows the high-level algorithm for retrieving relevant documents with \sys-FLAT.
We also visualize the \sys cache and workflow in \Cref{fig:cache_design} with an example when processing two subsequent similar query embeddings $ q_1 $ and $ q_2 $.
Each key in the cache corresponds to an embedding previously queried, while the associated value is a list of the top-$k$ nearest neighbors retrieved from the database during a previous query, where $k$ is a strictly positive integer constant picked at the start of the experiment.
The cache has a fixed capacity of $c$ key entries, which implies that, when full, an eviction policy is applied to make room for new entries (see \Cref{sec:eviction}).

When a user initiates a query, it is first converted into an embedding vector by an embedding model.
\sys is agnostic of the specific embedding model used but we assume that the embedding of the user query and the embeddings in the vector database are generated by the same embedding model, since embeddings from different models are not directly comparable.
The retriever (left in \Cref{fig:cache_design}) then forwards this query embedding, denoted as $ q_1 $, to the \sys cache {\color{Blue}\ding{202}}.
\sys first checks whether a similar query has been recently processed by iterating over each key-value pair ($k$, $v$) of cache entries (line \ref{algo1computedists} in \Cref{algo:search}).
If the best match is sufficiently close to the query, \ie, the distance between $q$ and $k$ is lower than some threshold $\tau$, the associated retrieval results are returned immediately (\hyperref[algo1minby]{lines 3-5}), thus bypassing the vector database.
Otherwise, the system proceeds with a standard database query (line \ref{algo1dbcall} in \Cref{algo:search}).
This is step {\color{Blue}\ding{203}} in \Cref{fig:cache_design}, where we perform a lookup with $ q_1 $ and the vector database.
The \sys cache is now updated with the resulting neighbor vector indices (in blue) from the database lookup {\color{Blue}\ding{204}}.
Since the number of cache entries might exceed the cache size, the eviction policy will remove a cache entry if necessary (\hyperref[algo1ifevict]{lines 7-8}).
The cache is now updated with the result obtained from the database (line \ref{algo1updatecache}).
Finally, the vector indices are returned to the retriever {\color{Blue}\ding{205}}.
We adopt the same distance function in \sys as the underlying vector database to ensure consistency between the caching mechanism and the retrieval process.

When another query embedding $ q_2 $ arrives {\color{Blue}\ding{206}} with a low distance to $ q_1$, \sys first checks if it is sufficiently similar to any stored query embeddings in the cache.
Suppose the distance between $ q_1 $ and $ q_2 $ is below the predefined similarity threshold $\tau$.
In that case, the cache returns the previously retrieved document indices associated with the closest matching query {\color{Blue}\ding{207}}, thus bypassing a lookup in the vector database.
We name this first approach \sys-FLAT as it scans the entire cache for every incoming query (line \ref{algo1computedists}) without using any data structure to guide it further.
Depending on the specifications of the cache and workload, \sys-FLAT can reduce retrieval latency and computational overhead, especially in workloads with strong spatial or temporal similarities.

\subsection{Scalability and \sys-LSH}
\label{sec:proximity_lsh}

While \sys-FLAT provides caching benefits with minimal complexity, its linear scan over all cached entries is a scalability bottleneck.
As the cache size increases, so does the computational cost of each query lookup, eventually increasing the end-to-end inference time.
This linear scan ensures that we find the closest match in the cache but also results in a lookup cost that is linearly dependent on $c$, the cache capacity.
For large caches, this cost becomes prohibitive.

To mitigate this scalability bottleneck, we introduce \sys-LSH, a variant of our approximate cache that supports scalable similarity-based lookups using \ac{LSH}, to enable the query to be redirected to a small bucket of entries that are the most likely to match it. This bounds the amount of query-key comparisons to be performed to the size of the bucket.
More specifically, we rely on random hyperplane \ac{LSH} \cite{10.1145/509907.509965}, a classical method for \ac{ANN} search in high-dimensional spaces.
The key idea is to compare each embedding to a fixed set of $ L $ randomly generated hyperplanes passing through the origin.
Each hyperplane defines a binary partition of space: a vector lies on one side or the other.
By repeating this for $L$ hyperplanes, we obtain an $L$-bit signature (a binary hash code) that serves as the key for the \ac{LSH} table.
Formally, given an embedding vector $q \in \mathbb{R}^d$, and $L$ random hyperplanes (stored as normal vectors ${r_1, r_2, \ldots, r_L}$), we compute its hash code $h(q)$ as:

$$h(q) : \{0,1\}^L = (q \cdot r_1 \geq 0, ... , q \cdot r_L \geq 0)$$\\

\begin{figure}[t]
    \centering
    \includegraphics{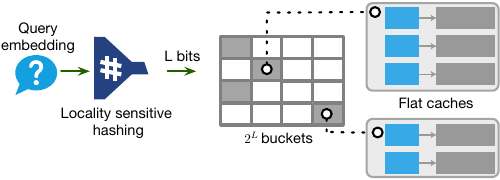}
    \caption{The workflow of \sys-LSH where each LSH bucket maps to a \sys-FLAT cache.}
    \label{fig:proximity_lsh}
    \Description[We show the workflow of Proximity-LSH where a query is hashed, mapped to a bucket and handled there]{We show the workflow of Proximity-LSH where a query is hashed, mapped to a bucket and handled there}
\end{figure}

In \sys-LSH, the hash code $h(q)$ determines the bucket to which an incoming query maps.
We show the workflow of \sys-LSH in \Cref{fig:proximity_lsh}.
Each bucket is a fixed-size \sys-FLAT cache with a capacity of $b = \num{20}$ entries, which we empirically find to strike a good balance between the hit rate and the execution latency per query (see \Cref{sec:bucketsize}).
When a query arrives, we compute its \ac{LSH} hash, identify its corresponding bucket, and perform a linear scan within only that bucket to check for similar embeddings.
If no similar entry is found, the query is forwarded to the vector database, and the bucket is updated accordingly.
Unlike when using a single flat cache, each bucket in \sys-LSH operates its own local eviction policy such as \ac{FIFO} or \ac{LRU}.
In particular, there is no global eviction policy that is shared across \sys-LSH caches as this would require additional data structures and state storage, resulting in additional compute and memory costs. Note that this makes \sys-LSH a $b$-way set-associative cache, where the set corresponding to a given entry is designated by $L$-bit \ac{LSH}.
We experimentally show in \Cref{sec:evaluation} that \ac{LSH}-based partitioning sustains high cache hit rates and accuracy.

Another benefit of using such buckets is the mitigation of false positives, \ie, cases where the similarity tolerance is large enough to consider a query and a cache line as matching, even though the two embeddings are actually semantically very different.
This mitigation arises because the tolerance only applies within the bucket to which the query maps, limiting potential false positives to that subset.
Due to the LSH property of grouping similar embeddings, cache lines in this bucket are more likely to be relevant.
Moreover, if a query is very infrequent, its corresponding bucket may be empty.
In such a case, false positives cannot occur, even though other buckets contain entries.
This contrasts with \sys-FLAT, where any cache line can potentially be a false positive depending on the global similarity threshold.

\textbf{Performance gains compared to \sys-FLAT.}
\sys-LSH yields a significant performance boost when dealing with caches with many entries.
In \sys-FLAT, every query is compared against all $c$ cached embeddings using the chosen distance metric, resulting in a per-query lookup cost of $\mathcal{O}(c \cdot d)$, where $d$ is the embedding dimensionality. This linear dependency on the cache capacity becomes noticeable in large caches, especially when $c$ reaches into the thousands, as discussed in \Cref{sec:lshvsflat}.

In contrast, \sys-LSH limits comparisons to a single fixed-size bucket.
Each query is hashed using $L$ random hyperplanes, at a cost of $\mathcal{O}(L \cdot d)$, and linearly compared only to the $b$ entries (\eg, $b=20$) in the selected bucket, with a comparison cost of $\mathcal{O}(b \cdot d)$ corresponding to the distance computations for each of the $b$ cache lines.
Both $L$ and $b$ are (small) constants, making the total cost of a lookup $\mathcal{O}(d)$.
Most importantly, it is independent of the total cache capacity. %

As an example, for $c = 10\,000 $ entries, $d = 768$, $b = 20$, and $L = 10$, a \sys-FLAT lookup performs $10\,000 \times 768 = 7.68$ million operations per query. In contrast, \sys-LSH performs only $(L + b) \cdot d = 30 \times 768 = 23\,040$ operations, a reduction of over 300$\times$ in per-query compute.
This performance gap widens with increasing cache size.
We show in \Cref{sec:evaluation} that this speedup does not come at the cost of the quality of the retrieved documents.

\subsection{\sys cache parameters and components}
\label{sec:cache_design}
We now describe the parameters and components of the \sys cache, and discuss their impact on performance.

\subsubsection{Cache capacity $ c $}
The cache has a capacity of $ c $ entries, which dictates the number of entries it will fill before starting to evict old entries (\ie, the number of rows in \Cref{fig:cache_design}).
This parameter poses a trade-off between the cache hit rate and the time it takes to scan the entire set of keys.
A larger cache increases the likelihood of cache hits, allowing the system to reuse previously retrieved documents more frequently and reducing the number of expensive vector database lookups.
However, increasing the cache size also incurs
\begin{enumerate*}[label=\emph{(\roman*)}]
\item computational overhead as the cache must be searched for similarity matches on each query and
\item memory overhead since additional key-value pairs need to be stored. %
\end{enumerate*}
For a small enough $c$, this computational overhead is manageable since the cache size is likely to be small compared to the full vector database.

\paragraph{Effective capacity in \sys-LSH}
Unlike \sys-FLAT, where the cache capacity $c$ is a flat global limit on the number of cached embeddings, \sys-LSH distributes this capacity across multiple buckets, each with a fixed maximum size $ b $ (\eg, 20 entries).
The total cache capacity of \sys-LSH is thus $c = 2^L \cdot b$, where $b$ is the per-bucket size and $L$ the number of random hyperplanes.

However, because \ac{LSH} partitions the embedding space based on the distribution of query vectors, in practice not all buckets receive traffic.
Some buckets remain unused, especially in workloads with skewed or clustered query distributions.
As a result, the actual number of stored entries is often much smaller than the theoretical maximum.
This sparsity leads to more efficient space usage:
we do not allocate memory for a given bucket until an entry is inserted. \sys-LSH allocates space only where needed, adapting naturally to the query distribution.
Our experiments in \Cref{sec:evaluation} show that with \sys-LSH only a fraction of the buckets are actively populated, and the memory footprint scales more gracefully with usage compared to flat caches of the same theoretical size.

\subsubsection{Eviction policy}
\label{sec:eviction}
When the cache reaches its maximum capacity, an eviction policy is required to determine which entry should be removed to make space for new ones.
The choice of policy can influence the hit rate of the cache, especially in workloads with strong temporal locality or repeated query patterns.
In \sys-FLAT, we support both \ac{FIFO} and \ac{LRU} eviction strategies.
\ac{FIFO} evicts the oldest inserted entry regardless of usage frequency.
It is straightforward to implement, incurs minimal overhead, and in our experiments, provides comparable accuracy to more sophisticated strategies in many settings.
However, \ac{FIFO} can underperform when there is strong temporal locality, \ie, when recent queries are more likely to be repeated.
In such cases, \ac{LRU} can yield higher hit rates by preferentially retaining recently accessed entries.
\ac{LRU} maintains a usage timestamp or recency ordering, evicting the entry that has gone unused the longest.
While slightly more expensive to manage, we find that \ac{LRU} remains efficient for cache sizes considered in \sys-FLAT and is particularly effective under bursty query traffic.
In \sys-LSH, eviction is managed separately within each individual bucket.

\subsubsection{Distance tolerance $\tau$}
We employ a fuzzy matching strategy based on a predefined similarity threshold $\tau$ to determine whether a cached entry can be used for an incoming query.
The choice of $\tau$ directly impacts recall and test accuracy, and is therefore a key parameter of \sys.
A low value of $\tau$ enforces stricter matching, ensuring that retrieved documents are highly relevant but potentially reducing the cache hit rate, thus limiting the benefits of caching.
We note that $\tau=0$ is equivalent to using a cache with exact matching.
Conversely, a higher value of $\tau$ increases cache utilization by accepting looser matches, improving retrieval speed at the potential cost of including less relevant documents.
In our experiments, $\tau$ is treated as a global constant, manually set at the start of each evaluation.
In general, determining the appropriate tolerance for high-dimensional approximate matching is a challenge that has no standard solution in the literature we reviewed.
Most notably, Frieder \etal~\cite{frieder2024caching} propose reusing the neighbor information to generate a dynamic tolerance per cache line.
We found that this still required some arbitrary hand-tuning.
In this paper, we leave $\tau$ as a hyperparameter to be optimized during deployment.
We explore the influence of this parameter on the cache performance in \Cref{sec:evaluation}.

\hyphenation{data-bases}
\subsubsection{The re-ranking factor $\rho$}
It is common practice for \ac{ANN} vector databases such as \diskann~\cite{jayaram2019diskann} and \textsc{FAISS}~\cite{douze2024faiss} to retrieve more neighbors than they actually return to the retriever~\cite{macdonald2021approximate}.
This over-fetching improves recall at the cost of increased latency.
We adopt a similar strategy in \sys.
Specifically, when querying the vector database (line \ref{algo1dbcall} in \Cref{algo:search}), we request a number of neighbors significantly exceeding the count required by the retriever.
This does not incur additional latency since the database inherently performs this broader search.
Upon a subsequent cache hit (line \ref{algo1earlyreturn} in \Cref{algo:search}), rather than immediately returning all associated vectors, we re-rank these vectors to prioritize and select only the ones that are most relevant to the current query.
We define the re-ranking factor $\rho \geq 1$ as the ratio between the number of vectors retrieved from the database and the number of vectors expected by the \ac{RAG} pipeline.

\section{Evaluation}
\label{sec:evaluation}

We implement both \sys-FLAT and \sys-LSH, and evaluate our approximate caching approach using two standard benchmarks: \Acf{MMLU} and \medrag.
Specifically, our experiments answer the following three questions:
\begin{enumerate}[label=\emph{(\roman*)}]
\item What is the impact of cache parameters such as cache capacity $c$, similarity tolerance $\tau$, \ac{LSH} hashing granularity $L$, and per-bucket capacity $b$ on the end-to-end test accuracy, k-recall, hit rate, and latency of \sys-FLAT and \sys-LSH (\Cref{sec:exp_impact_parameters})?
\item How does the cache occupancy of \sys-LSH change when varying the \ac{LSH} hashing granularity $L$ and similarity tolerance $\tau$ (\Cref{sec:exp_occupancy})?
\item How does the lookup time of \sys vary with cache occupancy and parameters such as cache capacity $c$, similarity tolerance $\tau$, and \ac{LSH} hashing granularity $L$? Furthermore, how well does \sys maintain hit rate and recall on large-scale datasets? (\Cref{sec:exp_scalability})
\end{enumerate}

\begin{figure*}
	\centering
	\includegraphics{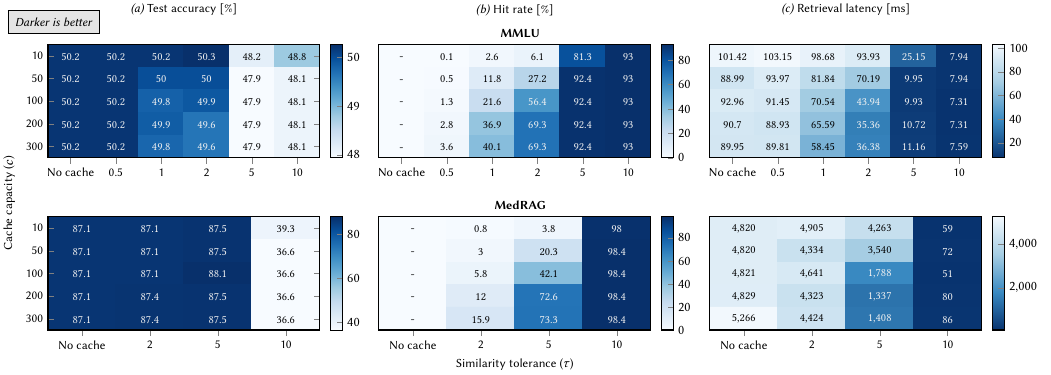}
	\caption{The test accuracy (a), cache hit rate (b), and latency of document retrieval (c) of \sys-FLAT, for different cache capacities and similarity tolerances, for the \mmlu (top) and \textsc{MedRAG} (bottom) benchmarks.}
	\Description[Heatmap grids showing caching performance for two datasets (MMLU and MedRAG).
    Rows represent cache capacities; columns represent similarity tolerances.
    Metrics shown are Test Accuracy, Hit Rate, and Retrieval Latency.
    Darker colors indicate better performance.]{Heatmap grids showing caching performance for two datasets (MMLU and MedRAG).
    Rows represent cache capacities; columns represent similarity tolerances.
    Metrics shown are Test Accuracy, Hit Rate, and Retrieval Latency.
    Darker colors indicate better performance. We show that there exists a sweet spot in terms of tolerance that enables large speedups at virtually no cost in accuracy.}
	\label{fig:heatmaps-lru}
\end{figure*}

\begin{figure*}
    \centering
    \includegraphics{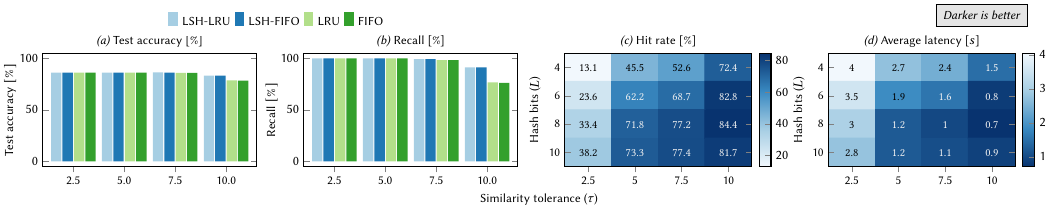}
    \caption{The test accuracy (a), recall (b), hit rate (c) and average latency (d), for different similarity tolerances, eviction policies, with and without LSH, under the \medrag-\textsc{Zipf} benchmark.
    	}
    \label{fig:e2eacc}
    \Description[Accuracy, recall, hit rate and latency plots for Medrad-Zipf]{Accuracy, recall, hit rate and latency plots for Medrad-Zipf. We show that accuracy remains constant with respect to tolerance up to tau = 7.5. Hit rates and latency improve with tolerance and cache size (measured by the number of hashing planes)}
\end{figure*}

\subsection{Implementation details}
We implement \sys in the Rust programming language and make its implementation available online.\footnote{https://github.com/sacs-epfl/proximity}
We use SIMD CPU instructions for all numerical computations (\eg the Euclidean distance compuation at line \ref{algo1computedists} of \cref{algo:search}).
Our implementation uses \textsc{portable-simd}\footnote{\url{https://github.com/rust-lang/portable-simd}}, an experimental extension to the Rust language that enables ISA-generic explicit SIMD operations.
While all LLVM target architectures are technically supported, we only considered modern x64 and ARM64 platforms for unit testing and performance evaluation.

We expose bindings from the Rust cache implementation to the Python machine learning pipeline using \textsc{PyO3}\footnote{\url{https://docs.rs/pyo3/latest/pyo3/}.} (Rust-side bindings generation) and \textsc{Maturin}\footnote{\url{https://www.maturin.rs}.} (Python-side package management).
These bindings streamline the integration of \sys into existing \ac{RAG} pipelines.

\subsection{Experimental setup}
To evaluate \sys, we adopt and modify two existing end-to-end \ac{RAG} workflows, \mmlu and  \textsc{MedRAG}, both introduced by previous work on \ac{LLM} question answering~\cite{adlakha2023evaluating,xiong-etal-2024-benchmarking}. %
In our setup, all vectors are stored in main memory without serialization to disk, which enables low-latency access to them.
We leverage the \textsc{FAISS} library~\cite{douze2024faiss} for efficient \ac{ANN} search.
For the \ac{LLM}, we use the open-source LLaMA 3.1 Instruct model~\cite{dubey2024llama}, which is optimized for instruction-following tasks.

\subsubsection{Document source.}
For the \mmlu benchmark, we use the \textsc{wiki\_dpr} dataset as a document source, which contains 21 million passages collected from Wikipedia.
The index used is \textsc{FAISS-HNSW}, a graph-based indexing structure optimized for fast and scalable \ac{ANN} search.
For \medrag, we use \pubmed as the document source, which contains 23.9 million medical publication snippets.
The associated vector database is served using \textsc{FAISS-Flat}.
For both \textsc{wiki\_dpr} and \pubmed, we embed each passage as a 768-dimensional vector.
We use the \textsc{MedCPT}~\cite{jin2023medcpt} and the \textsc{DPR}~\cite{karpukhin-etal-2020-dense} embedding models for the \medrag and \mmlu benchmarks, respectively.

\subsubsection{Query workload.}
We evaluate \sys by using a subset of the \mmlu and \textsc{PubMedQA} question datasets.

\textbf{\mmlu}
is a comprehensive benchmark designed to evaluate the knowledge and reasoning of \acp{LLM} across a wide array of subjects~\cite{hendrycks2020measuring}.
It encompasses approximately \num{16000} multiple-choice questions ranging between \num{57} topics, \eg, mathematics and history, with question difficulties ranging from elementary to advanced professional levels.
\mmlu is frequently used to evaluate the effectiveness of \ac{RAG}, and we leverage the subset of questions on the topic of econometrics, containing \num{131} total questions.
We specifically pick this subset because it highly benefits from \ac{RAG}. %

\textbf{\medrag}
is a benchmark for question answering in the biomedical domain, and we specifically focus on the \textsc{PubMedQA} question set that contains \num{500} questions and answers.
Similarly to \mmlu, we select at random \num{200} questions from \textsc{PubMedQA} to serve as user queries.
To simulate similarity, we generate four variants of each question by adding some small textual prefix to them and we randomize the order of the resulting \num{524} questions for \mmlu and \num{800} for \medrag.
These two datasets help us to showcase the performance of our cache implementation in the case where there is no strong bias in the queries used: every query appears four times in a slightly different format each time.
We denote these two datasets as the \emph{uniform} datasets, as they do not display bias in the queries.

\textbf{\medrag-\textsc{Zipf}.}
Finally, to showcase the capabilities of \sys in the context of strong, real-life-like bias, we create a synthetic third dataset based on \textsc{PubMedQA} by drawing \num{10000} queries from the original set of \num{500} \textsc{PubMedQA} questions with repetition, following a Zipf distribution with parameter 0.8, which we estimate to be a reasonable representation of real-world skews, as discussed in \Cref{sec:skew}.
Each time a given query appears, we rephrase it by asking an \ac{LLM}\footnote{gpt4o-2024-08-06 by OpenAI} to rephrase the question in a way that is syntactically different but semantically equivalent to the original.
We verify that each generated version of the question is unique across the entire dataset, and that the \ac{RAG} generative system indeed provides the same answer across all rephrasings when run without \sys.
The most frequently repeated question appears about \num{700} times in this dataset while the vast majority of original questions appear at most a few times.
Every original question appears at least once.
This simulates several users asking the same question with various wordings, in which case \sys should ideally only perform a single retrieval from the vector database, and rely on the cache for further queries.
As a worst-case scenario for caching, we do not simulate temporal locality of queries: all queries are statistically independent from each other, even though their distribution is severely biased towards some of the original \textsc{PubMedQA} questions.
We refer to this third dataset as the \medrag-\textsc{Zipf} dataset.

\subsubsection{Hardware.}
We launch our experiments in Docker containers, using 12 cores of an x64 Intel Xeon Gold 6240 CPU and \SI{300}{\giga\byte} of allocated RAM per container.

\subsubsection{Metrics.}
Our evaluation focuses on three performance metrics:
\begin{enumerate*}[label=\emph{(\roman*)}]
    \item The \emph{test accuracy} of the entire \ac{RAG} system, which is computed as the percentage of multiple-choice questions in our query workloads answered correctly by the \ac{LLM};
    \item The \emph{cache hit rate}, which is defined as the percentage of queries that find a sufficiently similar match in the cache;
    \item The \emph{retrieval latency}, which is the time required to retrieve the relevant documents, including both cache lookups and vector database queries where necessary.
\end{enumerate*}
To quantify how well the cache preserves the quality of retrieved context, we also measure the \emph{database $k$-recall}, defined as the fraction of the top-$k$ documents returned by the cache that are also among the top-$k$ results retrieved from the vector database for the same query.
This metric allows us to assess the extent to which cached results align with the true nearest neighbors, serving as a proxy for retrieval quality.
While the ultimate goal is to maintain end-to-end accuracy in the RAG pipeline, $k$-recall provides a more direct and inexpensive way to evaluate the effectiveness of approximate caching, independent of downstream \ac{LLM} variability.
To ensure statistical robustness, we run each experiment five times and with different random seeds.
We average all results. %

\subsection{The impact of cache parameters on test accuracy, recall, hit rate, and latency}
\label{sec:exp_impact_parameters}
We first examine the impact of the cache capacity $c$ and similarity tolerance $\tau$ on the three metrics described above.
We evaluate these metrics across different cache capacities $c \in$ \{\num{10}, \num{50}, \num{100}, \num{200}, \num{300}\} for both uniform benchmarks using the \ac{FIFO} cache eviction policy.
We experiment with tolerance levels $\tau \in \{0, 0.5, 1, 2, 5,10\} $ for \mmlu and $\tau \in \{0, 2, 5,10\} $ for \medrag. We perform no re-ranking ($\rho = 1$) for \mmlu and \medrag.
For the \medrag-\textsc{Zipf} dataset, we provide the accuracy and database recall for various cache eviction policies, with and without \ac{LSH} (see \Cref{sec:proximity_lsh}).
We also report the hit rate and average latency, as in the uniform datasets.
We use a re-ranking factor of $\rho = 4$ in experiments on \medrag-\textsc{Zipf}.
\Cref{fig:heatmaps-lru} shows the results on all uniform datasets and \Cref{fig:e2eacc} shows the results for \medrag-\textsc{Zipf}.

\subsubsection{Test accuracy}
\Cref{fig:heatmaps-lru}(a) shows the end-to-end \ac{RAG} accuracy for the \mmlu and \medrag benchmarks for different combinations of $ c $ and $ \tau $ and with \sys-FLAT.
The figure indicates that accuracy remains relatively stable across different combinations of $ c $ and $ \tau $, with values ranging between 47.9\% and 50.2\% for \mmlu (top row).
Test accuracy is slightly higher for low similarity tolerances $\tau = 0 $ (no cache) and $ \tau = 0.5$: approximately 50.2\%.
Increasing $\tau$ slightly degrades accuracy, bringing it closer to 48.1\%.
This is because a higher similarity tolerance increases the likelihood of including irrelevant documents in the \ac{LLM} prompt, negatively affecting accuracy.
We observe similar behavior in \medrag (bottom row), which shows a more pronounced accuracy drop between $\tau=5$ (88\%) and $\tau=10$ (37\%) for the same reason.
Increasing $c$ can lower accuracy, \eg, in \mmlu for $\tau = 1.0$, accuracy lowers from 50.2\% to 49.8\% when increasing $c$ from 10 to 300.
Interestingly, the highest observed accuracies of 50.3\% (for $\tau=3,c=10$ in \mmlu) and 88.1\% (for $\tau=5,c=100$ in \medrag) are achieved with caching.
This serendipitously occurs because the approximately retrieved documents prove more helpful on the \mmlu benchmark than the closest neighbors retrieved from the database without caching ($\tau=0$).
In both scenarios, the cache rarely decreases accuracy to the level of the \ac{LLM} without \ac{RAG} (48\% for \mmlu, 57\% for \medrag), except when the value of $\tau $ becomes too high (\eg, $\tau = 10$ for \medrag).

\Cref{fig:e2eacc}(a) shows the test accuracy when using the \medrag-\textsc{Zipf} benchmark, for different similarity tolerances, eviction policies, and with and without \ac{LSH}.
For $ \tau \in \{ 2.5, 5, 7.5 \}$, we observe comparable accuracies across the evaluated eviction policies and between \sys-FLAT and \sys-LSH.
For $ \tau = 10 $, we observe a notable degradation in accuracy for \sys-FLAT: from 85.7\% for \ac{LRU} with $ \tau = 2.5 $ to 77.4\% for \ac{LRU} with $ \tau = 10 $.
This same degradation is much less pronounced for \sys-LSH: from 85.8\% for \textsc{LSH-LRU} with $ \tau = 2.5 $ to 84.1\% for \textsc{LSH-LRU} with $ \tau = 10 $.
These results show that \sys-LSH is robust against higher values of $ \tau $.

\subsubsection{Recall}
\Cref{fig:e2eacc}(b) displays the k-recall for different similarity tolerances, eviction policies, and with and without \ac{LSH}, and when using the \medrag-\textsc{Zipf} benchmark.
This recall indicates the overlap between document indices returned by the cache and the document indices that the database \emph{would have returned} in the absence of caching.
We observe that the k-recall is virtually 100\% for tolerances below 7.5, showing that the cache is perfectly transparent in this regime: the retrieved documents by the cache are exactly the same as the ones that the database would have selected.
For $ \tau = 10 $, we observe a degradation in the k-recall, which translates to a similar degradation in the end-to-end accuracy of the model (see \Cref{fig:e2eacc} (a)).
This is consistent across all four cache configurations.
We note that this correlation between accuracy and k-recall can be leveraged to empirically fine-tune the tolerance hyperparameter $ \tau $ at a low cost, as k-recall can be evaluated without \ac{LLM} inference, which is compute-intensive.
Once the optimal tolerance has been found for k-recall, one can extrapolate that it is also optimal for end-to-end LLM accuracy.

\subsubsection{Cache hit rate}
\Cref{fig:heatmaps-lru}(b) shows the cache hit rate for different values of $ c $ and $\tau $ for both benchmarks and with \sys-FLAT.
Increasing $ \tau $ increases the hit rate.
For $\tau = 0$, there are no cache hits, as queries need to be equal to any previous query.
However, for $\tau \geq 5$, hit rates reach 93\% for \mmlu and 98.4\% for \medrag, demonstrating that higher tolerances allow the cache to serve most queries without contacting the database.
In this scenario, the cache prefers to serve possibly irrelevant data rather than contact the database.
Nevertheless, even with such high hit rates, there is only a minor decrease in accuracy for \mmlu.
Similarly, in \medrag, despite the larger drop in accuracy, a hit rate of 72.6\% ($\tau=5,c=200$) sustains an accuracy close to the upper bound.
Increasing the cache capacity significantly improves the hit rate, \ie, for \mmlu, $ \tau = 2 $, and when increasing $ c $ from 10 to 300, the hit rate increase from 6.1\% to 69.3\%.

Figure 7(c) further illustrates the cache hit rate as a function of the similarity tolerance $\tau$ and the number of LSH hash bits $L$, for \sys-LSH and with the skewed \medrag-\textsc{Zipf} workload.
For this experiments we use the \ac{LRU} eviction policy.
We observe that hit rates increase with $\tau$, confirming the trends seen \Cref{fig:heatmaps-lru}(b).
Importantly, even with small hash sizes in \sys-LSH (\eg, 4–6 bits, corresponding to 16 and 64 buckets, respectively), hit rates remain high for tolerances above 5.0, validating the effectiveness of \ac{LSH} in grouping semantically similar queries.

\subsubsection{Query latency}
\Cref{fig:heatmaps-lru}(c) shows the retrieval latency for different values of $ c $ and $ \tau $, for \sys-FLAT.
Latency reductions are significant for configurations with high cache hit rates.
For $\tau = 0$ (no cache) and $ c = 10 $, retrieval latency can be as high as \SI{101}{\milli\second} for \mmlu and \SI{4.8}{s} for \medrag.
Retrieval latency quickly decreases as $\tau$ increases, which aligns with the increase in hit rate; more queries are now answered with results from the cache.
Furthermore, increasing $ c $ also decreases retrieval latency, particularly for higher values of $ \tau $.
Finally, we remark that the speedup gains by \sys increase as the latency of vector database lookups increases.
While our implementation keeps all vector indices in main memory, other database implementations such as \diskann (partially) store indices on the disk, which further increases retrieval latency~\cite{jayaram2019diskann}.

\Cref{fig:e2eacc}(d) shows the average retrieval latency for \sys-LSH with the \ac{LRU} eviction policy.
Latency drops sharply as the similarity tolerance increases and the cache hit rate improves.
This improvement in average latency is again linearly increasing with the hit rate, as cache hits dismiss database calls entirely and cache misses are virtually zero-cost when compared to vector database calls.

\subsubsection{Bucket size}
\label{sec:bucketsize}
\begin{figure}
   \centering
   \includegraphics{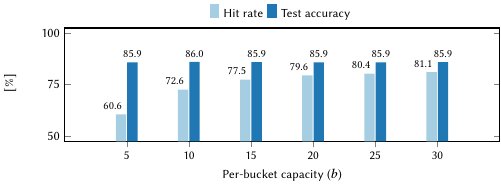}
   \caption{The test accuracy and hit rate for \sys-LSH as a function of the flat bucket size.}
   \label{fig:bucket_20}
   \Description[Exploring the optimal b parameter. 20 manages a good tradeoff between hit rate and per-bucket size]{Exploring the optimal b parameter. Several buckets sizes from 5-30 are explored. 20 manages a good tradeoff between hit rate and per-bucket size.}
\end{figure}

We now study the impact of the per-bucket capacity $b$ in \sys-\ac{LSH} on the hit rate and test accuracy.
Intuitively, larger bucket capacities increase the chance of finding a match among cached entries, thus improving the hit rate.
However, they also increase the number of entries to be scanned per lookup, resulting in higher per-query latency.

We evaluate this trade-off with the \medrag-\textsc{Zipf} dataset, using \sys-\ac{LSH} with $L=8$ hyperplanes, a tolerance parameter $\tau = 7.5$ and the \ac{LRU} eviction policy.
\Cref{fig:bucket_20} shows the test accuracy and hit rate for different bucket sizes $b$.
Our experimental results show that $b=20$ offers the best balance between hit rate and lookup cost.
Increasing $b$ from 5 to 20 substantially improves the hit rate from 60.6\% to 79.6\%, while the accuracy remains stable at approximately 85.9\%.
For $b > 20$, the hit rate quickly reaches a plateau (81.1\% at $b=30$), while accuracy remains unchanged.
Thus, larger bucket capacities yield negligible benefits in terms of hit rate or accuracy, yet impose higher scanning overhead.
We therefore fix $b=20$ in our experiments with \sys-\ac{LSH}.

\subsection{\sys-LSH cache occupancy}
\label{sec:exp_occupancy}
\begin{figure}
   \centering
   \includegraphics{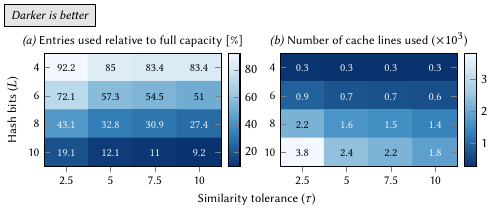}
   \caption{The relative (a) and absolute (b) cache occupancy after completion of the \medrag-\textsc{Zipf} workload when using the \sys-LSH cache with a \ac{LRU} eviction policy, for varying LSH hashing granularities and similarity tolerances.}
   \Description[Cache occupancy as a function of hash bits (L) and tolerance tau. Tolerance does not have a large impact. A larger count of hash bits decreases occupancy by a large margin, even though the amount of cache lines used increases.]{Cache occupancy as a function of hash bits (L) and tolerance tau. Tolerance does not have a large impact. A larger count of hash bits decreases occupancy by a large margin, even though the amount of cache lines used increases.}
   \label{fig:space_usage}
\end{figure}

We now determine the cache occupancy of \sys-LSH, under varying similarity tolerances $\tau$ and hashing granularity, expressed as the number of hash bits $ L $, after the \medrag-\textsc{Zipf} workload has completed. We note that in this experiment, the eviction policy (LRU or FIFO) is not relevant, as they both share the same memory footprint.
We show these results in \Cref{fig:space_usage}.
\Cref{fig:space_usage}(a) shows \emph{relative} usage, computed as the fraction of the theoretical maximum capacity, while \Cref{fig:space_usage}(b) reports the \emph{absolute} number of vectors stored in the cache.
The maximum capacity of \sys-LSH is $2^L \times 20$, where $L$ is the number of hash bits and 20 is the per-bucket limit.

Across all configurations, we observe a consistent trend: increasing the number of hash bits $ L $ significantly reduces relative cache utilization.
For instance, with $L = 4$ (\ie, a total of 16 buckets), the cache achieves over 92\% utilization with $\tau = 2.5$, filling 295 of the 320 entries.
In contrast, for $L = 10$ (1024 buckets), only 19.1\% of the cache is used at the same tolerance level, corresponding to \num{3920} occupied entries out of a maximum of \num{20480}.
This reflects the sparsity inherent to LSH-based caching: many buckets remain unused unless queries are uniformly distributed across the embedding space, which is not the case for our skewed workload.

Increasing the tolerance $\tau$ slightly \emph{reduces} both absolute and relative cache occupancy: for all configurations of $ L $, cache occupancy declines as $\tau$ increases from 2.5 to 10.0.
This is due to the increased hit rate at higher tolerances: more queries are matched to existing entries, thereby reducing the number of insertions required.
For example, at $L = 8$, usage decreases from \num{2209} entries (43.1\%) at $\tau = 2.5$ to \num{1402} entries (27.4\%) at $\tau = 10.0$.

We argue that this adaptive sparsity is a desirable property of \sys-LSH.
Unlike \sys-FLAT that gradually fills up to its maximum size, LSH-based caching only allocates memory where needed.
In practical deployments with skewed or clustered query distributions, this leads to significantly lower memory usage while preserving high cache effectiveness.
Consequently, \sys-\textsc{LSH} can be provisioned with a large theoretical capacity for collision avoidance, without incurring the cost of fully allocating all cache slots.

\begin{figure}[tb]
    \centering
    \includegraphics{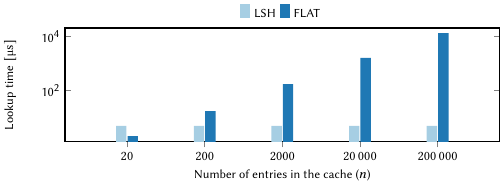}
    \caption{The cache lookup time as the number of cache entries increase, for \sys-FLAT (requiring a linear scan) and \sys-LSH. For both cache types we use the \ac{LRU} eviction policy.
        }
    \Description[The evolution of the lookup time in Proximity as a function of the amount of lines in the cache. LSH has constant runtime, whereas FLAT has a linear dependency on the amount of lines.]{The evolution of the lookup time in Proximity as a function of the amount of lines in the cache. LSH has constant runtime, whereas FLAT has a linear dependency on the amount of lines.}
    \label{fig:lat_comp}
\end{figure}

\begin{figure}[h]
    \centering
    \includegraphics{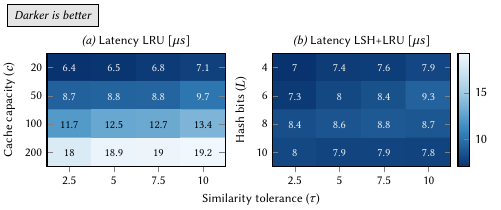}
    \caption{The cache lookup time of \medrag-\textsc{Zipf} queries sent to \sys-FLAT (a) and \sys-LSH (b) for varying cache capacities and similarity tolerances.}
    \Description[The evolution of the lookup time in Proximity as a function of the cache size and tolerance. LSH has constant runtime, whereas FLAT has a linear dependency on the amount of lines.]{The evolution of the lookup time in Proximity as a function of the cache size and tolerance. LSH has constant runtime, whereas FLAT has a linear dependency on the amount of lines.}
    \label{fig:lat_comp_two}
\end{figure}

\subsection{\sys scalability}
\label{sec:exp_scalability}
Finally, we evaluate the scalability of \sys by analyzing the lookup times for different cache occupancies, and when varying cache parameters. We also provide experimental insights on deploying \sys on larger query sets by analyzing the recall and hit rate on the \tripclick dataset.

\subsubsection{Lookup times for increasing cache occupancies}
\label{sec:lshvsflat}
We now evaluate the cache lookup times by \sys-FLAT and \sys-LSH by measuring the time spent per query as the cache size increases, under the \ac{LRU} eviction policy.
We vary the number of entries that are stored in the cache when the query arrives, denoted by $ n $ and visualize the results in \Cref{fig:lat_comp}.
\Cref{fig:lat_comp} (dark blue bars) shows the per-query latency of the \sys-FLAT implementation.
The lookup time increases nearly linearly with the cache capacity $ c $ (this holds regardless of the similarity tolerance $\tau$, as the amount of computation to be done is independent of $\tau$).
Specifically, this increase is from \SI{2.0}{\micro\second} for $ n = 20 $ to \SI{13.0}{\milli\second} for $ n = $ \num{200000}.
This is consistent with the algorithmic complexity of \sys-FLAT, which performs a full scan of the cache and computes the distance between the incoming query and every stored key.
Even at reasonable cache sizes, this linear scan incurs an overhead that can become a noticeable proportion of the retrieval time budget.
In contrast, \Cref{fig:lat_comp} (light blue bars) displays the scaling capabilities of \sys-LSH in terms of the amount of vectors that are stored in the cache when the user query arrives at the cache. We use $L = 8$ hashing planes.
The lookup time remains constant as $ n $ increases: for all evaluated values of $ n $ we observe a lookup time of merely \SI{4.8}{\micro\second}.
This highlights the excellent scalability of the \sys-LSH cache.

\subsubsection{Lookup times for varying cache parameters}
We run \sys with the \medrag-\textsc{Zipf} benchmark and analyze the cache lookup times of queries sent to both \sys-FLAT and \sys-LSH ($L=8$), for varying cache capacities and similarity tolerances (\Cref{fig:lat_comp_two}).
Unlike the measurements in \Cref{fig:e2eacc}(d), which capture the full end-to-end retrieval time including database access, for this experiment we only consider the time spent performing cache lookups.
\Cref{fig:lat_comp_two}(a) shows these results for \sys-FLAT and highlights that increasing $ c $ while fixing $ \tau $ increases the cache lookup time.
This is because there are more cache entries and thus more computational requirements to complete the lookup.
A similar trend is visible when increasing $ \tau $ when keeping $ c $ fixed.
This is because for increasing values of $ \tau $ we are left with more candidates after the linear scan completes (line \ref{algo1computedists} in \Cref{algo:search}) and therefore require more time in determining the best candidate (line \ref{algo1minby} in \Cref{algo:search}).
\Cref{fig:lat_comp_two}(b) shows the latency of \sys-LSH for varying tolerance and LSH hashing granularity.
Here, the lookup time remains relatively stable across hashing granularity and similarity tolerances.
The time per query depends only on the per-bucket capacity, and not on the total number of cached entries.
This confirms our theoretical expectation that \ac{LSH}-based cache lookups have constant-time complexity with respect to both total cache size and cache tolerances.

\subsubsection{Large query dataset}
\label{sec:tripclickcache}
\begin{figure}
	\centering
	\includegraphics{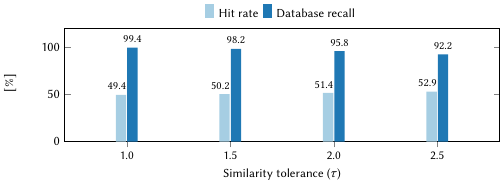}
	\caption{Hit rate and database k-recall of \sys-\textsc{LSH} on the \tripclick dataset and with the \ac{LRU} eviction policy, while varying the similarity tolerance $\tau$.}
    \Description[Hit rate and recall of Proximity on TripClick as a function of the tolerance tau. Hit rate remains relatively stable at 50\%, recall slowly degrades. Optimal tau is 1.0.]{Hit rate and recall of Proximity on TripClick as a function of the tolerance tau. Hit rate remains relatively stable at 50\%, recall slowly degrades. Optimal tau is 1.0.}
	\label{fig:tripclickcache}
\end{figure}

To demonstrate the effectiveness of \sys on a large-scale query dataset, we use \tripclick as the query set (see \Cref{sec:skew}), \pubmed as the document database, and \diskann as the vector database.
\tripclick contains approximately 5.2 million user queries and we send queries to the system in the same order as they appear in the original dataset.
We evaluate both the cache hit rate and the database recall by comparing the vectors returned by \sys-\textsc{LSH} ($L=8$) against those retrieved directly from the database for the same queries.
\Cref{fig:tripclickcache} shows how the cache hit rate and database recall behave for different values of the similarity tolerance $\tau$ when using the \ac{LRU} eviction policy.
\sys-\textsc{LSH} maintains a stable cache hit rate around $50\%$ across all similarity tolerance levels $\tau$.
For $\tau=1.0$ and $ \tau = 1.5 $, the cache effectively reduces the number of retrieval calls by approximately $50\%$ while preserving near-perfect k-recall.
However, k-recall decreases as $\tau$ increases, \eg, from 99.4\% for $ \tau=1.0 $ to 92.2\% for $\tau = 2.5$.
This is because a higher similarity tolerance permits the cache to match a larger number of cache lines.
These findings further highlight that \sys can achieve significant latency reductions while maintaining high retrieval accuracy even when using workloads with a large number of queries and with real-world spatial and temporal locality.

\subsection{Experimental conclusion}
Our findings demonstrate that \sys effectively reduces retrieval latency while maintaining competitive accuracy.
This makes approximate caching a viable optimization for RAG pipelines in scenarios where queries exhibit spatial and temporal similarity.
In practical deployments, however, tuning the tolerance and cache capacity hyperparameters based on workload characteristics will be critical to balancing performance and accuracy.
 
\section{Related work}
\label{sec:related}
\textbf{Improving \ac{RAG} latency.}
Various strategies have been proposed to decrease the retrieval latency of  \ac{RAG}.
Zhu et al. propose Sparse RAG, an approach that encodes retrieved documents in parallel, thereby reducing delays associated with sequential processing~\cite{zhu2024accelerating}.
Sparse RAG reduces overhead of the \ac{LLM} encoding stage.
\textsc{RAGServe} is a system that dynamically adjusts parameters, such as the number of retrieved documents, for each query~\cite{ray2024ragserve}. 
The system balances response quality and latency by jointly scheduling queries and adapting configurations based on individual query requirements.
\textsc{PipeRAG} integrates pipeline parallelism and flexible retrieval intervals to accelerate \ac{RAG} systems through concurrent retrieval and generation processes~\cite{jiang2024piperag}.
\textsc{RAGCache} is a multilevel dynamic caching system that organizes intermediate states of retrieved knowledge into a hierarchical structure, caching them across GPU and host memory to reduce overhead~\cite{jin2024ragcache}.
\textsc{TurboRAG} reduces the latency of the prefill phase by caching and reusing \ac{LLM} key-value caches~\cite{lu2024turborag}.
Cache-Augmented Generation is a method that preloads all relevant documents into a language model's extended context and precomputes key-value caches, thus bypassing real-time retrieval during inference~\cite{chan2024don}.
Speculative RAG improves accuracy and reduces latency by using a smaller \ac{LLM} to generate multiple drafts in parallel from subsets of retrieved documents~\cite{wang2024speculative}.
The above systems optimize different aspects of the \ac{RAG} workflow and many of them are complementary to \sys.

\textbf{Similarity caching.}
Beyond \ac{RAG}, caching mechanisms have long been studied to improve efficiency in information retrieval systems~\cite{alonso1990data}. In particular, similarity-based caching techniques aim to reduce retrieval latency and improve system throughput by exploiting patterns in query semantics and distribution~\cite{chierichetti2009similarity,pandey2009nearest}. These approaches typically leverage the observation that similar queries tend to retrieve similar results, enabling cache reuse even in the absence of exact query matches.

Such techniques have found applications across a variety of domains, including approximate image retrieval~\cite{falchi2012similarity}, content distribution networks~\cite{nakamura2024analysis}, and recommendation systems~\cite{sermpezis2018soft}. In these contexts, approximate caching is often implemented using distance metrics or clustering methods in embedding spaces to identify and reuse semantically close queries.

However, to the best of our knowledge, these approaches have not been applied in the context of question answering, where approximate nearest neighbor search (NNS) is used not merely to retrieve similar content, but to support factual or task-oriented responses via downstream LLM inference. This makes correctness and latency trade-offs more sensitive and requires a finer-grained control over recall and relevance.

In this work, we extend approximate caching to this domain by introducing \sys, a system designed specifically for low-latency, large-scale \ac{RAG}.
Our design incorporates key scalability optimizations, including locality-sensitive hashing to reduce lookup complexity and explicit SIMD instructions to accelerate distance computations.
These architectural decisions enable our system to operate efficiently under realistic workloads and tight latency budgets, making it, to our knowledge, the first practical application of approximate similarity caching in \ac{RAG}-based question answering pipelines.

\textbf{Answer caching.}
Finally, Regmi et al.~\cite{regmi2024gpt} propose a semantic caching strategy tailored for \ac{LLM} applications, where responses to previous user queries are stored and reused when semantically similar queries are detected. In their system, incoming user queries are embedded into a high-dimensional space, and the cache returns a previously generated LLM response if a sufficiently similar query has already been seen. Their approach bypasses both document retrieval and answer generation entirely on cache hits, leading to substantial latency reductions.

While effective in reducing end-to-end response time, this technique assumes that semantically close queries always require identical responses, which may not hold in practice. This is especially the case when subtle distinctions between queries are relevant. In contrast, \sys focuses on optimizing the \ac{RAG} document retrieval phase by caching references to retrieved documents rather than final answers. This allows the \ac{LLM} to regenerate responses tailored to the precise query phrasing, preserving potentially important nuances that may be lost at the embedding level. As a result, our method provides greater flexibility and better preserves fidelity to the user's intent, while still reducing the computational overhead of expensive nearest-neighbor lookups. 
\section{Conclusion}
\label{sec:conclusion}

We introduced \sys, a novel caching mechanism designed to enhance the efficiency of \ac{RAG} systems.
Our approach significantly reduces retrieval latency while maintaining retrieval accuracy by leveraging spatial and temporal similarities in user queries to \acp{LLM}.
Instead of treating each query as an independent event, \sys caches results from previous queries and reuses them when similar queries appear.
This caching reduces the computational load on the underlying vector database and decreases the end-to-end latency of the overall \ac{RAG} pipeline.
We also improve the scalability of query lookup times by designing \sys-LSH, a cache that uses \acf{LSH} to dramatically speed up the process of determining similar queries in the cache.
Our evaluation with the \mmlu and \medrag benchmarks demonstrates that both \sys-FLAT and \sys-LSH provide substantial performance gains in scenarios where users repeatedly query related topics.
We conclude that our approximate caching strategy effectively optimizes \ac{RAG} pipelines, particularly in workloads with similar query patterns.

\begin{acks}
This work has been funded by the Swiss National Science Foundation, under the project ``FRIDAY: Frugal, Privacy-Aware and Practical Decentralized Learning'', SNSF grant number \href{https://data.snf.ch/grants/grant/10001796}{10001796}.
We are also grateful to Micha\l{} Stawarz for helpful discussions and for his recommendations regarding relevant literature.
\end{acks}

\bibliographystyle{ACM-Reference-Format}
\bibliography{main.bib}

\end{document}